\documentclass{svmult}

\usepackage[latin1]{inputenc}
\usepackage[english]{babel}
\usepackage{amsmath,amssymb}
\usepackage{eufrak}
\usepackage[mathscr]{eucal}
\usepackage{color}
\usepackage[dvips]{graphicx}
\usepackage{fancyhdr}
\usepackage{cite} 
\newcommand{\bbE}{\mathbb{E}}\newcommand{\bbF}{\mathbb{F}}

\newcommand{\bbN}{\mathbb{N}}
\newcommand{\bbP}{\mathbb{P}}\newcommand{\bbR}{\mathbb{R}}

\newcommand{\bbZ}{\mathbb{Z}}

\newcommand{\frakF}{\mathfrak{F}}

\newcommand{\frakP}{\mathfrak{P}}

\newcommand{\frakX}{\mathfrak{X}}

\newcommand{\tonda}[1]{\left( #1 \right)}
\newcommand{\quadra}[1]{\left[ #1 \right]}
\newcommand{\graffa}[1]{\left\{ #1 \right\}}
\newcommand{\norma}[1]{\left\| #1 \right\|}

\newcommand{\modulo}[1]{\left| #1 \right|}
\newcommand{\doubleindex}[2]{
\textrm{\tiny$
\begin{array}{c}
#1 \\
#2
\end{array}
$}}

\newtheorem{Lem}{Lemma}[chapter]

\newtheorem{Teo}[Lem]{Theorem}
\newenvironment{teo}[1][]{\begin{Teo}\begin{normalfont}\emph{#1 }}
{\end{normalfont}\finenu\end{Teo}}
\newtheorem{Cor}[Lem]{Corollary}
\newenvironment{cor}[1][]{\begin{Cor}\begin{normalfont}\emph{#1 }}
{\end{normalfont}\finenu\end{Cor}}
\newtheorem{Pro}[Lem]{Proposition}
\newenvironment{pro}[1][]{\begin{Pro}\begin{normalfont}\emph{#1 }}
{\end{normalfont}\finenu\end{Pro}}
\newtheorem{Defi}[Lem]{Definition}
\newenvironment{defi}[1][]{\begin{Defi}\begin{normalfont}\emph{#1 }}
{\end{normalfont}\fine\end{Defi}}
\newtheorem{Oss}[Lem]{Remark}
\newenvironment{oss}[1][]{\begin{Oss}\begin{normalfont}\emph{#1 }}
{\end{normalfont}\fineoss\end{Oss}}
\newtheorem{Es}[Lem]{Example}
\newenvironment{es}[1][]{\begin{Es}\begin{normalfont}\emph{#1 }}
{\end{normalfont}\fineoss\end{Es}}
\newcounter{HPcont}[section]
\renewcommand{\theHPcont}{(A-\arabic{HPcont})}
\newenvironment{assume}{
\refstepcounter{HPcont} \myitem[\textbf{\theHPcont}]} {\par}

\newenvironment{myproof}[1][]{\par\noindent\textbf{Proof{#1}. }}{\finedim\par}

\newcommand{\fine}{}
\newcommand{\finenu}{}
\newcommand{\finedim}{\hfill$\blacksquare$}
\newcommand{\fineoss}{}

\newcounter{numerazione}
\newenvironment{mynumerate}{\setcounter{numerazione}{0}}{}
\newcommand{\mynumber}[1][\refstepcounter{numerazione}(\arabic{numerazione}) ]{\par\noindent #1 }

\newcommand{\myitem}[1][$\bullet$]{\par\noindent #1 $\ \ $}

\newcounter{Bnumero}[Lem]

\newcommand{\Banach}{\frakX} 

\newcommand{\Ball}[2]{B(#1,#2)}

\newcommand{\Mink}{\oplus} 
\newcommand{\Minkowski}{+} 
\newcommand{\HausDist}{\delta_H} 

\newcommand{\Parts}[1]{\frakP'} 
\newcommand{\eParts}[1]{\frakP} 

\newcommand{\Closed}[2][]{\bbF'_{#1}} 

\newcommand{\eClosed}[2][]{\bbF_{#1}} 

\newcommand{\closure}[1]{\overline{#1}}
\newcommand{\interior}[1]{\textrm{Int}\ #1}

\newcommand{\cmpt}{K}
\newcommand{\cpt}{K}

\newcommand{\cut}[2]{{#1}_{#2}}
\newcommand{\speed}{G}

\newcommand{\Dspeed}{G}

\newcommand{\nucleation}{B}

\newcommand{\tildeSpeed}[2]{\widehat{\speed}^{#1}_{#2}}
\newcommand{\add}[2][W]{\tonda{\partial^{\Minkowski \cpt}_{#1} #2}}

\newcommand{\salgebra}{\frakF} 

\newcommand{\prob}[1][]{\bbP_{#1}}

\newcommand{\leb}{\mu_\lambda}

\newcommand{\expec}{\bbE}
\newcommand{\condExpec}[2][\cdot]{\expec\quadra{\left.{#1}\right|{#2}}}

\newcommand{\racs}{{RaCS}}

\newcommand{\supKzero}{\sup_{\doubleindex{\cmpt\in\eClosed[k]{\Banach}}{\cmpt\subseteq \cmpt_0}} }

\let\oldmarginpar\marginpar
\renewcommand\marginpar[1]{\-\oldmarginpar[\raggedleft\scriptsize #1]%
{\raggedright\scriptsize #1}}
\newcommand{\notamargine}[1]{\marginpar{ }}

\begin{document}

\title{Statistical aspects of birth--and--growth stochastic processes}
\author{Giacomo Aletti, Enea G. Bongiorno, Vincenzo Capasso}
\institute{Department of Mathematics, University of Milan,\\ via Saldini 50, 10133 Milan Italy\\
\texttt{giacomo.aletti@mat.unimi.it} \\
\texttt{bongio@mat.unimi.it} \\
\texttt{vincenzo.capasso@mat.unimi.it}}
%
%
\maketitle

\abstract{The paper considers a particular family of set--valued 
stochastic processes modeling birth--and--growth processes. The proposed setting allows us to investigate the nucleation and the growth processes. A decomposition theorem is established to characterize the nucleation and the growth. As a consequence, different consistent set--valued estimators are studied for growth process. Moreover, the nucleation process is studied via the hitting function, and a consistent estimator of the nucleation hitting function is derived.}

\bibliographystyle{plain}

\section*{Introduction}

Nucleation and growth processes arise in several natural and technological applications (cf. \cite{cap03,cap03a} and the references therein) such as, for example, solidification and phase--transition of materials, semiconductor crystal growth, biomineralization, and DNA replication (cf., e.g., \cite{her:jun:bec:ben02}).
During the years, several authors studied stochastic spatial processes (cf. \cite{cre91, sto:ken:mec95, mol97} 
and references therein) nevertheless they essentially consider static approaches modeling real phenomenons. For what concerns the dynamical point of view, a parametric \emph{birth--and--growth process} was studied in \cite{molJ92,molJ95}. A birth--and--growth process is a \racs{} family given by $ \Theta_t=\bigcup_{n:T_n\le t} \Theta_{T_n}^t(X_n)$, for $t\in\bbR_{+} $, where $\Theta^t_{T_n}\tonda{X_n}$ is the \racs{} obtained as the evolution up to time $t > T_n$ of the germ born at (random) time $T_n$ in (random) location $X_n$, according to some growth model.
An analytical approach is often used to model birth--and--growth process, in particular it is assumed that the growth of a spherical nucleus of infinitesimal radius is driven according to a non--negative normal velocity, i.e.\ for every instant $t$, a border point of the crystal $x\in\partial \Theta_t$ ``grows'' along the outwards normal unit (e.g. \cite{fro:tho87, bur:cap:piz06, bur:cap:mic07, chi04}). In view of the chosen framework, different parametric and non--parametric estimations are proposed over the years (cf. \cite{molJ:sor94, mol:chi00, erh01, cap03a, cap:vil05, ale:saa08, chi:mol:qui03} and references therein).
Note that the existence of the outwards normal vector imposes a regularity condition on $\partial \Theta_t$ (and also on the nucleation process: it cannot be a point process).
\\%
On the other hand, it is well known that random sets are particular cases of fuzzy sets. Now, in the class of all convex fuzzy sets having compact support, Doob--type decomposition for sub- and super--martingales was studied (e.g. \cite{fei:wu04,fen:zhu02,fen00,ter04}). Nevertheless, a more general case (than the convex one) has not yet been considered; surely, in order to do this, the first easiest step is to consider decomposition for random set--valued processes.
After which, the step forward, to be considered in a following paper, can be to generalize results of this paper to birth--and--growth fuzzy set--value processes.

This paper is an attempt to offer an original approach based on a purely geometric stochastic point of view in order to avoid regularity assumptions describing birth--and--growth processes.
The pioneer work \cite{mic:pat:vil05} studies a growth model for a single convex crystal based on Minkowski sum, whilst in \cite{ale:bon:cap08}, the authors derive a computationally tractable mathematical model of such processes that emphasizes the geometric growth of objects without regularity assumptions on the boundary of crystals.
Here, in view of this approach, we introduce different set--valued parametric estimators of the rate of growth of the process. They arise naturally from a decomposition via Minkowski sum and they are consistent as the observation window expands to the whole space.
On the other hand, keeping in mind that distributions of random closed sets are determined by Choquet capacity functionals and that the nucleation process cannot be observed directly, the paper provides an estimation procedures of the hitting function of the nucleation process.

The article is organized as follows. Section \ref{sec: preliminary results} contains preliminary properties. Section \ref{sec: birthGrowth model} introduces a birth--and--growth model for random closed sets as the combination of two set--valued processes (nucleation and growth respectively). Further, a decomposition theorem is established to characterize the nucleation and the growth. Section \ref{sec: estimators of G} studies different estimators of the growth process and correspondent consistent properties are proved. In Section \ref{sec: capacity of nucleation}, the nucleation process is studied via the hitting function, and a consistent estimator of the nucleation hitting function is derived.

\section{Preliminary results}\label{sec: preliminary results}

Let $\bbN$, $\bbZ$, $\bbR$, $\bbR_{+}$ be the sets of all non--negative integer, integer, real and non--negative real numbers respectively, and let $\Banach=\bbR^d$. Let $\eClosed{\Banach}$ be the family of all closed subsets of $\Banach$ and $\Closed{\Banach} = \eClosed{\Banach}\setminus\{\emptyset\}$.
The suffixes $b$, $k$ and $c$ denote boundedness, compactness and convexity properties respectively (e.g. $\eClosed[kc]{\Banach}$ denotes the family of all compact convex subsets of $\Banach$).
\\%
For all $A,B\subseteq \Banach$ and $\alpha\in\bbR_{+}$, let us define
$$
\begin{array}{rll}
A\Minkowski B =& \graffa{a+b:a\in A,\ b\in B}=\bigcup_{b\in B}
b\Minkowski A,\qquad& \textrm{(Minkowski Sum)},
\\%
\alpha\cdot A=&\alpha A =\graffa{\alpha a: a\in A}, &
\textrm{(Scalar Product)},
\\%
A\ominus B =& \tonda{A^C \Minkowski B}^C=\bigcap_{b\in
B}b\Minkowski A,\qquad& \textrm{(Minkowski Subtraction)},
\\%
\check{A}=& \graffa{-a:a\in A},  & \textrm{(Symmetric Set)},
\end{array}
$$
where $A^C=\graffa{x\in\Banach: x\not\in A}$ is the complementary set of $A$, $x \Minkowski A$ means $\{x\} \Minkowski A$ (i.e.\ $A$ translate by vector $x$), and, by definition, $\emptyset\Minkowski A=\emptyset = \alpha \emptyset$. It is well known that $\Minkowski$ is a commutative and associative operation with a neutral element but, in general, $A\subseteq \Banach$ does not admit opposite (cf. \cite{rad52,kei:rot92}) and $\ominus$ is not the inverse operation of $\Minkowski$. The following relations are useful in the sequel (see \cite{ser84}): for every $A,B,C\subseteq \Banach$
$$
\begin{array}{c}
(A\cup B)\Minkowski C = (A\Minkowski C)\cup(B\Minkowski C),
\\%
\textrm{if }B\subseteq C, \quad A\Minkowski B\subseteq
A\Minkowski C,
\\%
(A\ominus B)\Minkowski \check{B} \subseteq A \quad\textrm{and}\quad%
(A\Minkowski B)\ominus \check{B} \supseteq A,%
\\%
(A \cup B)\ominus C \supseteq (A\ominus C) \cup (B\ominus C).
\end{array}
$$
In the following, we shall work with closed sets. In general, if $A,B\in\eClosed{\Banach}$ then $A\Minkowski B$ does not belong to $\eClosed{\Banach}$ (e.g., in $\Banach=\bbR$ let $A=\graffa{n+1/n : n> 1}$ and $B=\bbZ$, then $\graffa{1/n=\tonda{n+1/n}+(-n)}\subset A\Minkowski B$ and $1/n\downarrow 0$, but $0\not\in A\Minkowski B$). In view of this fact, we define $A\Mink B =\closure{A\Minkowski B}$ where $\closure{(\cdot)}$ denotes the closure in $\Banach$. It can be proved that, if $A\in\eClosed{\Banach}$ and $B\in\eClosed[k]{\Banach}$ then $A\Minkowski B\in\eClosed{\Banach}$ (see \cite{ser84}).

For any $A,B\in\Closed{\Banach}$ the \emph{Hausdorff distance} (or \emph{metric}) is defined by
$$
\HausDist(A,B) = \max\graffa{\sup_{a\in A}\inf_{b\in
B}\norma{a-b}_\Banach,
\sup_{b\in B} \inf_{a\in A}\norma{a-b}_\Banach}.%
$$

A random closed set (\racs{}) is a map $X$ defined on a probability space $(\Omega,\salgebra,\prob)$ with values in $\eClosed{\Banach}$ such that $\graffa{\omega\in\Omega : X(\omega)\cap K\neq \emptyset}$ is measurable for each compact set $K$ in $\Banach$.
%
%
It can be proved (see \cite{li:ogu:kre02}) that, if $X,X_1,X_2$ are \racs{} and if $\xi$ is a measurable real--valued function, then $X_1\oplus X_2$, $X_1\ominus X_2$, $\xi X$ and $(\interior{X})^C$ are \racs{}.
Moreover, if $\graffa{X_n}_{n\in\bbN}$ is a sequence of \racs{} then $X=\closure{\bigcup_{n\in\bbN} X_n}$ is so.

Let $X$ be a \racs{}, then $T_X(\cmpt)=\prob(X\cap \cmpt \neq \emptyset)$, for all $\cmpt\in \eClosed[k]{\Banach}$, is its \emph{hitting function} (or \emph{Choquet capacity functional}). The well known Matheron Theorem states that, the probability law $\prob[X]$ of any \racs{} $X$ is uniquely determined by its hitting function (see \cite{mat75}) and hence by $Q_X(\cmpt) = 1-T_X(\cmpt)$.
\begin{oss}[{(See \cite{mol93}.)}]\label{oss: capacity functional of XUY}\label{oss: hitting function of X+Y}
If both $X$ and $Y$ are \racs{}, then, for every $\cmpt\in\eClosed[k]{\Banach}$,
$$
T_{X\Mink Y}(\cmpt) = \expec\quadra{ \condExpec[T_{X}\tonda{\cmpt\Mink
\check{Y}}]{Y} }.
$$
Moreover, if $X,Y$ are independent, then, for every $\cmpt\in\eClosed[k]{\Banach}$,
$$
T_{X\cup Y}\tonda{\cmpt} = T_X\tonda{\cmpt} + T_Y\tonda{\cmpt} -
T_X\tonda{\cmpt} T_Y\tonda{\cmpt}.
$$
\end{oss}
A \racs{} $X$ is \emph{stationary} if the probability laws of $X$ and $X\Minkowski v$ coincide for every $v\in\Banach$. Thus, the hitting function of a stationary \racs{} clearly is invariant up to translation $ T_{X}(\cmpt)=T_{X}(\cmpt\Minkowski {v}) $ for each $\cmpt\in\eClosed[k]{\Banach}$ and any $v\in \Banach$.
\\%
A stationary \racs{} $X$ is \emph{ergodic}, if, for all $\cmpt_1,\cmpt_2\in \eClosed{\Banach}$,
$$
\frac{1}{\modulo{W_n}}\int_{W_n} Q_X((\cmpt_1\Minkowski v) \cup \cmpt_2) dv
\rightarrow Q_X(\cmpt_1)Q_X(\cmpt_2),\qquad \textrm{as} \qquad n\rightarrow
\infty;
$$
where $\graffa{W_n}_{n\in\bbN}$ is a \emph{convex averaging sequence of sets} in $\Banach$ (see \cite{dal:ver03}), i.e. each $\graffa{W_n}$ is convex and compact, $W_n\subset W_{n+1}$ for all $n\in\bbN$ and
$$
\sup\graffa{r\ge 0 : \Ball{x}{r}\subset W_n\textrm{ for some }x\in
W_n}\uparrow \infty, \qquad \textrm{as}\qquad n\rightarrow\infty.
$$
\begin{pro}\label{pro: X+Y stationary}
Let $X,Y$ be \racs{} with $Y\in\Closed[k]{\Banach}$ a.s. and $X$ stationary, then $X\Minkowski Y$ is a stationary \racs{}. Moreover, if $X$ is ergodic, then $X\Minkowski Y$ is so.
\end{pro}
\begin{myproof}
Let $Z=X\Minkowski Y$, it is a \racs{}. Note that
\begin{eqnarray*}
T_Z(\cmpt) &=& \expec\quadra{ \condExpec[T_{X}\tonda{\cmpt\Minkowski
\check{Y}}]{Y} } = \expec\quadra{ \condExpec[T_{X}\tonda{\cmpt\Minkowski
\check{Y}\Minkowski v}]{Y} } = T_Z(\cmpt\Minkowski v),
\end{eqnarray*}
for every $\cmpt\in\eClosed[k]{\Banach}$ and $v\in\Banach$, then $Z=X\Minkowski Y$ is stationary. Further, let us suppose that $X$ is ergodic, then, by Tonelli's Theorem and by dominated convergence theorem, we obtain
\begin{eqnarray*}
\int_{W_n} \frac{Q_Z((\cmpt_1\Minkowski v) \cup \cmpt_2) }{\modulo{W_n}}dv &=&
\expec\quadra{\condExpec[\frac{1}{\modulo{W_n}} \int_{W_n}
Q_{X}(((\cmpt_1\Minkowski v)\cup \cmpt_2) \Minkowski \check{Y}) dv ]{Y}}
\\%
&\rightarrow& \expec \quadra{\condExpec[ Q_{X}(\cmpt_1 \Minkowski \check{Y})
Q_{X}(\cmpt_2 \Minkowski \check{Y}) ]{Y}}
\\%
&=&Q_{Z}(\cmpt_1) Q_{Z}(\cmpt_2),
\end{eqnarray*}
for every $\cmpt_1,\cmpt_2\in\eClosed[k]{\Banach}$. Hence $X\Minkowski Y$ is ergodic.
\end{myproof}

\section{A Birth--and--Growth process}\label{sec: birthGrowth model}

Let $(\Omega,\salgebra, \graffa{\salgebra_n}_{n\in\bbN}, \prob)$ be a filtered probability space with the usual properties. Let $\graffa{\nucleation_n: n\ge 0}$ and $\graffa{\Dspeed_n: n\ge 1}$ be two families of \racs{} such that $\nucleation_n$ is $\salgebra_{n}$--measurable and $\Dspeed_n$ is $\salgebra_{n-1}$--measurable. These processes represent the \emph{birth} (or \emph{nucleation}) \emph{process} and the \emph{growth process} respectively. Thus, let us define recursively a birth--and--growth process $\Theta=\graffa{\Theta_n:n\ge 0}$ by
\begin{equation}\label{eq: discrete set process}
\Theta_n= \left\{
\begin{array}{ll}
(\Theta_{n-1}\Mink \Dspeed_{n} ) \cup \nucleation_n, & n\ge 1,\\
\nucleation_0, & n=0.
\end{array}\right.
\end{equation}
Roughly speaking, Equation \eqref{eq: discrete set process} means that $\Theta_n$ is the enlargement of $\Theta_{n-1}$ due to a Minkowski \emph{growth} $\Dspeed_n$ while  \emph{nucleation} $\nucleation_n$ occurs.
Without loss of generality let us consider the following assumption.
\begin{assume}\label{hp:0_in_G}
For every $n\ge 1$, $0\in \Dspeed_n$.
\end{assume}
Note that, Assumption \ref{hp:0_in_G} is equivalent to $\Theta_{n-1} \subseteq \Theta_n$.
\\%
In \cite{ale:bon:cap08}, the authors derive \eqref{eq: discrete set process} from a continuous time birth--and--growth process; here, in order to make inference, the discrete time case is sufficient.
Indeed, a sample of a birth--and--growth process is usually a time sequence of pictures that represent process $\Theta$ at different temporal step; namely $\Theta_{n-1}$, $\Theta_{n}$.
Thus, in view of \eqref{eq: discrete set process}, it is interesting to investigate $\{\Dspeed_n\}$ and $\graffa{\nucleation_n}$; in particular, we shall estimate the maximal growth $\Dspeed_n$ and the capacity functional of $\nucleation_n$. For the sake of simplicity, $Y$, $X$, $\Dspeed$ and $\nucleation$ will denote \racs{} $\Theta_n$, $\Theta_{n-1}$, $\Dspeed_n$ and $\nucleation_n$ respectively (then $X\subseteq Y$).
Thus, let us consider the following general definition.
\begin{defi}
Let $Y$, $X$ be \racs{} with $X\subseteq Y$. A \emph{$X$--decomposition of $Y$} is a couple of \racs{} $(\Dspeed,\nucleation)$ for which
\notamargine{$\Mink$ e non $+$ dato che non so se $\Dspeed$ è limitato}
\begin{equation}\label{eq: decomposition}
Y=(X \Mink \Dspeed) \cup \nucleation.
\end{equation}
\end{defi}
Note that, since we can consider $\tonda{\Dspeed, \nucleation}= \tonda{\graffa{0},Y}$, there always exists a $X$--decomposition of $Y$. It can happen that $\Dspeed$ and $\nucleation$ in \eqref{eq: decomposition} are not unique. As example, let $Y=[0,1]$ and $X=\{0\}$, then both $(\Dspeed_1,\nucleation _1)=(Y,Y)$ and $(\Dspeed_2,\nucleation _2)=(X,Y)$ satisfy \eqref{eq: decomposition}.
As a consequence, since we can not distinguish between two different decompositions, we shall choose a maximal one according to the following proposition.
\begin{pro}[(See \cite{ser84})]\label{pro: speed is maximum}
Let $Y$, $X$ be \racs{} with $X\subseteq Y$. Then
\begin{equation}\label{eq: G = g : g+B in A}
\speed=Y\ominus \check{X} = \graffa{g\in\Banach: g\Minkowski X \subseteq Y}.
\end{equation}
is the greatest \racs{}, with respect to set inclusion, such that $(X \Mink \speed)\subseteq Y$.
\end{pro}
\begin{cor}\label{cor: G' and G produce same effects}
The couple $(\speed=Y\ominus \check{X}, \nucleation= Y\cap\closure{(X\Mink \speed)^C})$ is the max-min $X$--decomposition of $Y$. As a consequence, $(\speed, \nucleation)$ is a $X$--decomposition of $Y$ and for any other $X$--decomposition
of $Y$, say $(\speed',\nucleation')$, then $\Dspeed'\subseteq\speed$ and $\nucleation'\supseteq \nucleation$.
\end{cor}
In other words, if $X,\Dspeed',\nucleation'$ are \racs{} and $Y=(X\Mink \Dspeed')\cup \nucleation' $, then $ \speed = Y\ominus \check{X} \supseteq \Dspeed'$ and $Y=(X\Mink \speed)\cup \nucleation' $.
\\%
Let $\Theta$ be as in \eqref{eq: discrete set process}. From now on, $\Dspeed_n$ denotes $\Theta_n\ominus \check{\Theta}_{n-1}$ that, as a consequence of Assumption \ref{hp:0_in_G}, contains the origin. Moreover, we shall suppose
\begin{assume}\label{hp:G_in_K}
There exists $\cpt\in\Closed[b]{\Banach}$ such that $G_n=\Theta_n\ominus \check{\Theta}_{n-1} \subseteq \cpt$ for every $n\in\bbN$.
\end{assume}
\begin{assume}\label{hp:B-Theta=0}
For every $n\ge 1$, $\tonda{\nucleation_n\ominus \check{\Theta}_{n-1}}=\emptyset$
almost surely.
\end{assume}
Roughly speaking, Assumption \ref{hp:G_in_K} means that process $\Theta$ does not grow too ``fast'', whilst Assumption \ref{hp:B-Theta=0} means that it cannot born something that, up to a translation, is larger (or equal) than what there already exists.

\notamargine{In realtà un'osservazione analoga è già nei preliminari. Penso che valga la pena ricordarla.}Let us remark that Assumption \ref{hp:G_in_K} implies $\graffa{\Dspeed_n}\subset \Closed[k]{\Banach}$ and $X\Mink \Dspeed_n = X\Minkowski \Dspeed_n$, for any \racs{} $X$.

\section{Estimators of $\Dspeed$}\label{sec: estimators of G}

On the one hand Proposition \ref{pro: speed is maximum} gives a theoretical formula for $\speed$, but, on the other hand, in practical cases, data are bounded by some observation window and edge effects may cause problems. Hence, as the standard statistical scheme for spatial processes (e.g. \cite{mol97}) suggests, we wonder if there exists a consistent estimator of $\speed$ as the observation window expands to the whole space $\Banach$.
\begin{pro}\label{pro: classical ---> our definition windows}
If $\graffa{W_i}_{i\in\bbN}\subset \Closed[ck]{\Banach}$ is a convex averaging sequence of sets, then, for any $\cpt\in\Closed[k]{\Banach}$, $\Banach = \bigcup_{i\in\bbN} W_i\ominus \check{\cpt} $. In this case, we say that $\graffa{W_i}_{i\in\bbN}$ expands to $\Banach$ and we shall write $W_i\uparrow \Banach$.
\end{pro}
\begin{myproof}
At first note that $\Banach = \bigcup_{i\in\bbN} \interior{W_i} $ and for any $i\in\bbN$, $W_i\subseteq W_{i+1}$.
\\%
Let $x\in\Banach$ and $\cpt\in\Closed[k]{\Banach}$. Note that, $x\Minkowski \cpt\in\Closed[k]{\Banach}$ is a compact set. Then there exists a finite family of indices $I\subset\bbN$ such that, if $N=\max I$, then
$$
x\Minkowski\cpt \subseteq \bigcup_{j\in I} \interior{W_{j}} =
\interior{W_{N}}.
$$
Hence, we have that $ x \in \interior{W_{N}} \ominus \check{\cpt} \subseteq W_{N} \ominus \check{\cpt}$, i.e., for any $x\in\Banach$, there exists $n_0\in\bbN$ such that $x\in W_{n_0} \ominus \check{\cpt}$.
\end{myproof}
Let $W\in\graffa{W_i}_{i\in\bbN}$ be an observation window and let us denote by $\cut{Y}{W}$ and $\cut{X}{W}$, the (random) observation of $Y$ and $X$ through $W$, i.e.\ $Y\cap W$ and $X\cap W$ respectively. Let us consider the estimator of $\speed$ given by the maximal $\cut{X}{W}$--decomposition of $\cut{Y}{W}$:
\begin{equation}\label{eq: wrong estimate}
\tildeSpeed{}{W}=\tonda{\cut{Y}{W}\ominus \cut{\check{X}}{W}}
\end{equation}
so that $\cut{X}{W} \Mink \tildeSpeed{}{W}\subseteq \cut{Y}{W}\subseteq W$. Notice that, whenever $Y$ and $X$ are bounded, then there exists $W_j\in\graffa{W_i}_{i\in\bbN}$ such that $Y\subseteq W_{j}$ and $\check{X}\subseteq W_{j}$, hence $\tildeSpeed{}{W_j}=Y\ominus \check{X} = \speed$. In other words, on the set $\graffa{\omega\in\Omega : X(\omega), Y(\omega) \textrm{ bounded}}$, the estimator \eqref{eq: wrong estimate} is consistent
$$
\tildeSpeed{}{W_i}(Y,X | Y,X\textrm{ bounded}) \rightarrow \speed,
\quad\qquad \textrm{as }W_i\uparrow\Banach;
$$
otherwise, as we already said, if $Y$ and $X$ are unbounded, edge
effects may cause problems and the estimator \eqref{eq: wrong
estimate} is, in general, not consistent as we discussed in the
following example.
\begin{es}
Let $\Banach=\bbR^2$, let us consider $X=\tonda{ \{x=0\} \cup \{y=0\} }$ and $Y=X\Minkowski \Ball{0}{1}$ where $\Ball{0}{1}$ is the closed unit ball centered in the origin. Surely $X\subset Y$, and they are unbounded. Note that $Y=(X\Minkowski \Dspeed)$ for any $\Dspeed$ such that $\tonda{\{0\}\times [-1,1] \cup [-1,1]\times \{0\} }\subseteq \Dspeed \subseteq \Ball{0}{1}$. On the other hand, by Proposition \ref{pro: speed is maximum}, there exists a unique $\speed$ that is the greatest set, with respect to set inclusion; in this case $\speed=[-1,1]\times[-1,1]$.
\\%
Let us suppose $0\in W_0$ and let $W\in\graffa{W_i}_{i\in\bbN}$, then, by Equation \eqref{eq: wrong estimate}, the estimator of $\speed$ is $\tildeSpeed{}{W}=\{0\}\neq\speed$. This is an edge effect due to the fact that, for every $\speed'$ with $\graffa{0}\subset \speed'\subseteq\speed$, it holds $\tonda{\cut{X}{W} \Minkowski \speed'} \cap W^C \neq \emptyset$ and then $\cut{X}{W} \Minkowski \speed'\not\subseteq \cut{Y}{W}$ that does not agree with Proposition \ref{pro: speed is maximum}.
\end{es}
Edge effects can be reduced by considering the following estimators
of $\speed$\notamargine{specificare ruolo di K. Bisogna usare K giusto altrimenti insorgono problemi al bordo.}
\begin{eqnarray}
\label{eq: G1(W,K)K} \tildeSpeed{1}{W} &=& \tonda{\cut{Y}{W}
\ominus \cut{\check{X}}{W\ominus\check{\cpt}}} \cap \cpt,
\\
\label{eq: G2(W,K)K} \tildeSpeed{2}{W} &=& \tonda{\quadra{
\cut{Y}{W} \cup \add{\cut{X}{W}} } \ominus
\cut{\check{X}}{W}} \cap \cpt;
\end{eqnarray}
where $\cpt$ is given in Assumption \ref{hp:G_in_K} and where $\add{\cut{X}{W}}=\closure{(\cut{X}{W} \Minkowski \cpt) \setminus W}$. The role of $\cpt$ will be clarified in Proposition \ref{pro: properties tildeSpeed} where it guarantees the monotonicity of $\tildeSpeed{1}{W}$.
Note that, estimators \eqref{eq: G1(W,K)K} \eqref{eq: G2(W,K)K} are bounded (i.e.\ compact) \racs{}, moreover, if $Y$ and $X$ are bounded, then $\tildeSpeed{1}{W_j}$, $\tildeSpeed{2}{W_j}$ eventually coincide with the estimator \eqref{eq: wrong estimate}; i.e.\ there exists $n_0$ such that for all $j\ge n_0$, $\tildeSpeed{}{W_j}= \tildeSpeed{1}{W_j}= \tildeSpeed{2}{W_j}=\speed$.

Let us explain how $\tildeSpeed{1}{W}$ and $\tildeSpeed{2}{W}$ work.
Estimator $\tildeSpeed{1}{W}$ is obtained by reducing the information given by $X$ to the smaller window $W\ominus\check{\cpt}$, whilst $Y$ is observed in $W$. Then $\tildeSpeed{1}{W}$ is the greatest subset of $\cpt$, with respect to set inclusion, such that $\cut{X}{W\ominus \check{\cpt}}\Minkowski \tildeSpeed{1}{W} \subseteq \cut{Y}{W}$ (see Proposition \ref{pro: speed is maximum}).
Estimator $\tildeSpeed{2}{W}$ is obtained by observing $X$ in $W$ (and not $W\ominus \check{\cpt}$), whilst $Y$ is increased (at least) by $\add{\cut{X}{W}}$, that is the greatest possible set of growth for $X$ outside of the observed window $W$. Then $\tildeSpeed{2}{W}$ is the greatest subset of $\cpt$, with respect to set inclusion, such that $(\cut{X}{W}\Minkowski \tildeSpeed{2}{W} )\cap W \subseteq \cut{Y}{W}$, or, alternatively, $\cut{X}{W}\Minkowski \tildeSpeed{2}{W} \subseteq \cut{Y}{W'}$, where $\cut{Y}{W'}= \cut{Y}{W}\cup \add{\cut{X}{W}}$ (see Proposition \ref{pro: speed is  maximum}).

Note that by definition of Minkowski Subtraction
\begin{eqnarray*}
\tildeSpeed{1}{W} =& \bigcap_{x\in\cut{X}{W\ominus\check{\cpt}}}&
x\Minkowski\tonda{(-x\Minkowski\cpt)\cap \cut{Y}{W}},
\\%
\tildeSpeed{2}{W} =& \bigcap_{x\in\cut{X}{W}} & x\Minkowski
\tonda{(-x\Minkowski\cpt) \cap Y_{W'}};
\end{eqnarray*}
i.e.\ every $x\in \cut{X}{W\ominus\check{\cpt}}$ (resp.
$x\in\cut{X}{W}$) ``grows'' at most as $(-x\Minkowski\cpt)\cap
\cut{Y}{W}$ (resp. $(-x\Minkowski\cpt)\cap \cut{Y}{W'}$).

Now, we are ready to show the consistency property of
$\tildeSpeed{1}{W_i}$ and $\tildeSpeed{2}{W}$. In particular,
Proposition \ref{pro: properties tildeSpeed} proves that
$\tildeSpeed{1}{W_i}$ decreases, with respect to set inclusion, to
the theoretical $\speed$, whenever $W_i$ expands to the whole space
($W_i \uparrow \Banach$). Proposition \ref{pro: G2 consistent as W
expand to X} proves that, for every $W\in\Closed{\Banach}$,
$\tildeSpeed{2}{W}$ is a better estimator than $\tildeSpeed{1}{W}$
and hence it is a consistent estimator of $\speed$.
\begin{pro}\label{pro: properties tildeSpeed}
Let $Y$, $X$ be \racs{}, let $0\in\speed=Y\ominus\check{X}\subseteq \cpt$. The following statements hold for $\tildeSpeed{1}{W}$:
\begin{mynumerate}
\mynumber $\speed\subseteq \tildeSpeed{1}{W}$ for every $W$;
\mynumber $\tildeSpeed{1}{W_{2}}\subseteq \tildeSpeed{1}{W_1}$ if
$W_2\supseteq W_1$;
\mynumber If $W_i\uparrow \Banach$, then $ \bigcap_{i\in\bbN}
\tildeSpeed{1}{W_i} = \speed$. Moreover,
\begin{equation}\label{eq: teo.properties tildeSpeed}
\lim_{i\to\infty} \HausDist(\tildeSpeed{1}{W_i} , \speed ) = 0.
\end{equation}
\end{mynumerate}
\end{pro}
\begin{myproof}
\begin{mynumerate}
\mynumber Since $0\in\cpt$, $\bigcap_{k\in\cpt}-k\Minkowski W=W\ominus\check{\cpt}\subseteq W$ and then $\cut{X}{W\ominus\check{\cpt}}\subseteq W$.
Let $g\in\speed$, then $g\Minkowski X\subseteq Y$. Since $g\in \cpt$, last inclusion still holds when $X$ and $Y$ are substituted by $\cut{X}{W\ominus\check{\cpt}}$ and $\cut{Y}{W}$ respectively: $g\Minkowski\cut{X}{W\ominus\check{\cpt}}\subseteq \cut{Y}{W}$.
Thus $g\in\tildeSpeed{1}{W}$ follows by Definition \eqref{eq:
G1(W,K)K} and Proposition \ref{pro: speed is maximum}.
\mynumber In order to obtain $\tildeSpeed{1}{W_{2}}\subseteq
\tildeSpeed{1}{W_1}$, it is sufficient to prove that
\begin{equation}\label{eq: XW1+GW2 < YW1}
\cut{X}{W_1\ominus\check{\cpt}} \Minkowski \tildeSpeed{1}{W_{2}} \subseteq \cut{Y}{W_{1}}
\end{equation}
since $\tildeSpeed{1}{W_1}$ is the greatest set, with respect to
set inclusion, for which the inclusion \eqref{eq: XW1+GW2 < YW1}
holds.
In fact, $W_1\ominus \check{\cpt} \subseteq \tonda{W_1 \ominus \check{\cpt}}\Minkowski \cpt \subseteq W_1\subseteq W_{2}$, then $\cut{X}{W_1\ominus\check{\cpt}}\subseteq \cut{X}{W_{2}}$.
Let $x\in \cut{X}{W_1\ominus\check{\cpt}} = X\cap
\tonda{W_1\ominus\check{\cpt}}$, then $x\in\cut{X}{W_2}$. By definition of $\tildeSpeed{1}{W_{2}}$, we have
$$
x\Minkowski\tildeSpeed{1}{W_{2}}\subseteq \cut{Y}{W_{2}} \subseteq
Y.
$$
On the other hand, since $x\in W_{1}\ominus \check{\cpt}$ and $\tildeSpeed{1}{W_{2}}\subseteq \cpt$, we have
$$
x\Minkowski\tildeSpeed{1}{W_{2}}\subseteq
\tonda{W_1\ominus\check{\cpt}} \Minkowski \cpt \subseteq W_{1};
$$
i.e.\ $x\Minkowski\tildeSpeed{1}{W_{2}}$ is included both in $Y$
and in $W_{1}$.
\mynumber Since $\speed\subseteq \bigcap_{i\in\bbN}\tildeSpeed{1}{W_i}$,
it remains to prove that
$$
\bigcap_{i\in\bbN}\tildeSpeed{1}{W_i}\subseteq \speed;
$$
i.e.\ if $g\in \tildeSpeed{1}{W_i}$ for each $i\in\bbN$, then
$g\in\speed$.
Take $g\in\bigcap_{i\in\bbN}\tildeSpeed{1}{W_i}$. By definition of
$\tildeSpeed{1}{W_1}$, we have
\begin{equation}\label{eq: g+b in Y^h}
g+x\in Y \textrm{ for all }
x\in\cut{X}{W_i\ominus\check{\cpt}}\textrm{ and }\forall i\in\bbN.
\end{equation}
By contradiction, assume $g\not\in \speed$. Then $g\Minkowski
X\not\subseteq Y$, i.e.\ there exists $\overline{x}\in X$ such
that $\tonda{g + \overline{x}}\not \in Y$. On the one hand, Proposition \ref{pro: classical ---> our definition windows} implies that there exists $j\in\bbN$ such that $\overline{x}\in W_j\ominus\check{\cpt}$. On the other hand, Equation \eqref{eq: g+b in Y^h} implies $g+ \overline{x}\in Y$ which is a contradiction.
Thus Theorem 1.1.18 in \cite{li:ogu:kre02} implies \eqref{eq: teo.properties tildeSpeed}.
\end{mynumerate}
\end{myproof}

\begin{pro}\label{pro: G2 consistent as W expand to X}
For every $W\in\Closed{\Banach}$, $\speed \subseteq
\tildeSpeed{2}{W}\subseteq \tildeSpeed{1}{W}$.
\end{pro}
\begin{myproof}
Let us divide the proof in two parts; in the first one we prove
that $\tildeSpeed{2}{W}\subseteq \tildeSpeed{1}{W}$, in the second
one that $\speed\subseteq \tildeSpeed{2}{W}$.
Let $g\in\tildeSpeed{2}{W}$ and $x\in X_{W\ominus\check{\cpt}}$.
Since $\tildeSpeed{2}{W}\subseteq\cpt$, we have
\begin{equation}\label{eq: 1pro.G2 consistent as W expand to X}
x+g\in \tonda{W\ominus\check{\cpt}}\Minkowski\tildeSpeed{2}{W}
\subseteq \tonda{W\ominus\check{\cpt}}\Minkowski \cpt \subseteq W;
\end{equation}
where we use properties of monotonicity of the Minkwoski
Subtraction and Sum. Moreover, by definition of
$\tildeSpeed{2}{W}$,
$$
x+g\in \cut{Y}{W},\qquad \textrm{or}\qquad x+g\in
\add{\cut{X}{W}}\subseteq W^C.
$$
By \eqref{eq: 1pro.G2 consistent as W expand to X}, $x+g\in \cut{Y}{W}$. The arbitrary choice of $x\in\cut{X}{W\ominus\check{\cpt}}$ completes the first part of the proof.
For the second part, let $g\in\speed$ and $x\in\cut{X}{W}$. By definition of $\speed$, $x+g\in Y$. We have two cases:
\myitem[-] $x+g\in W$, and therefore $x+g\in \cut{Y}{W}$,
\myitem[-] $x+g\not\in W$. Since $x\in \cut{X}{W}$, $
x+g\in \tonda{\cut{X}{W}\Minkowski \speed}\setminus W\subseteq
\add{\cut{X}{W}}.
$
\end{myproof}
\begin{cor}\label{cor: G2 consistent}
$\tildeSpeed{2}{W}$ is consistent (i.e.
$\tildeSpeed{2}{W}\downarrow\speed$ whenever $W\uparrow\Banach$).
\end{cor}

\begin{figure}[!htb]
\begin{center}
\includegraphics[clip, width=0.45\textwidth]{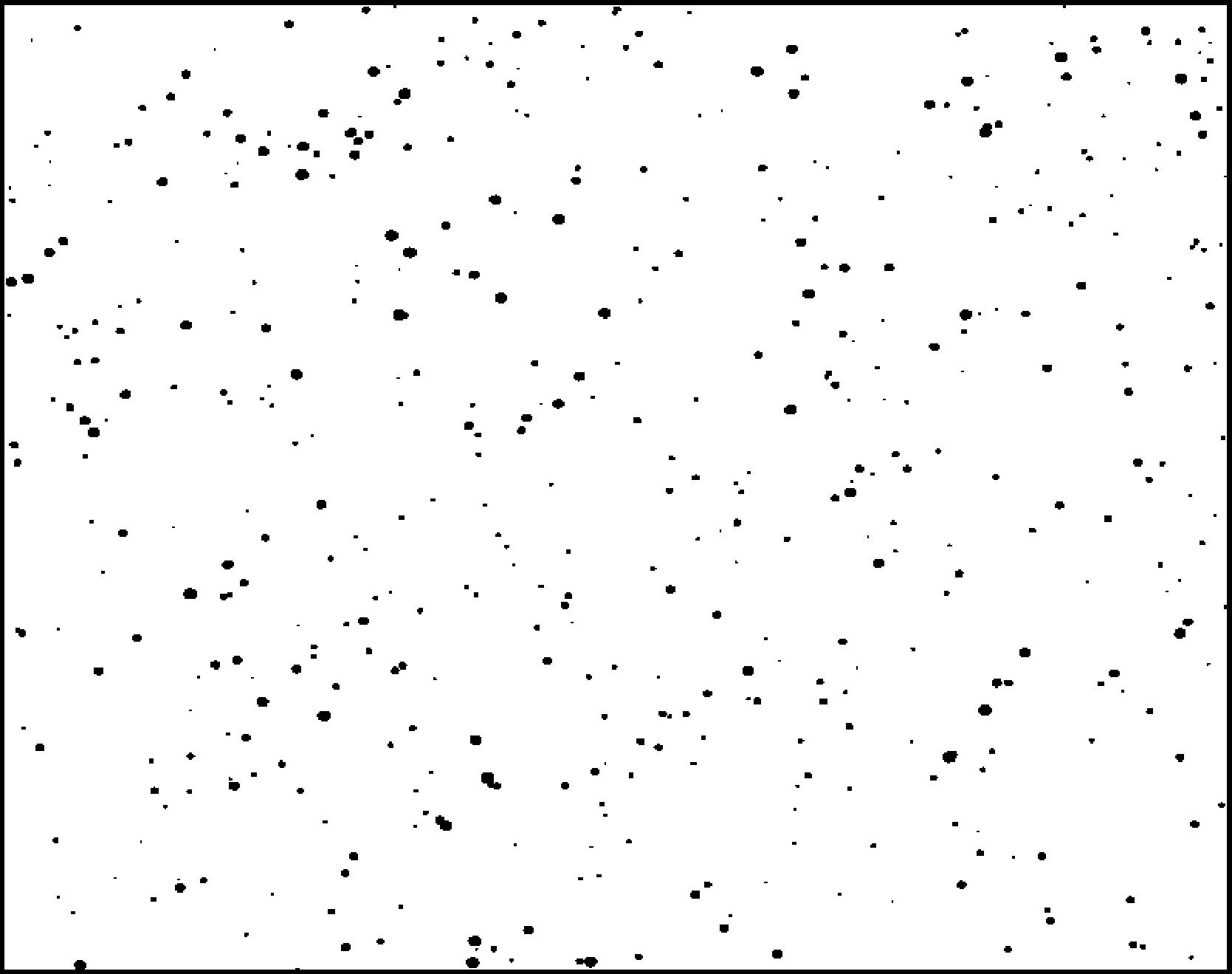}
\includegraphics[clip, width=0.45\textwidth]{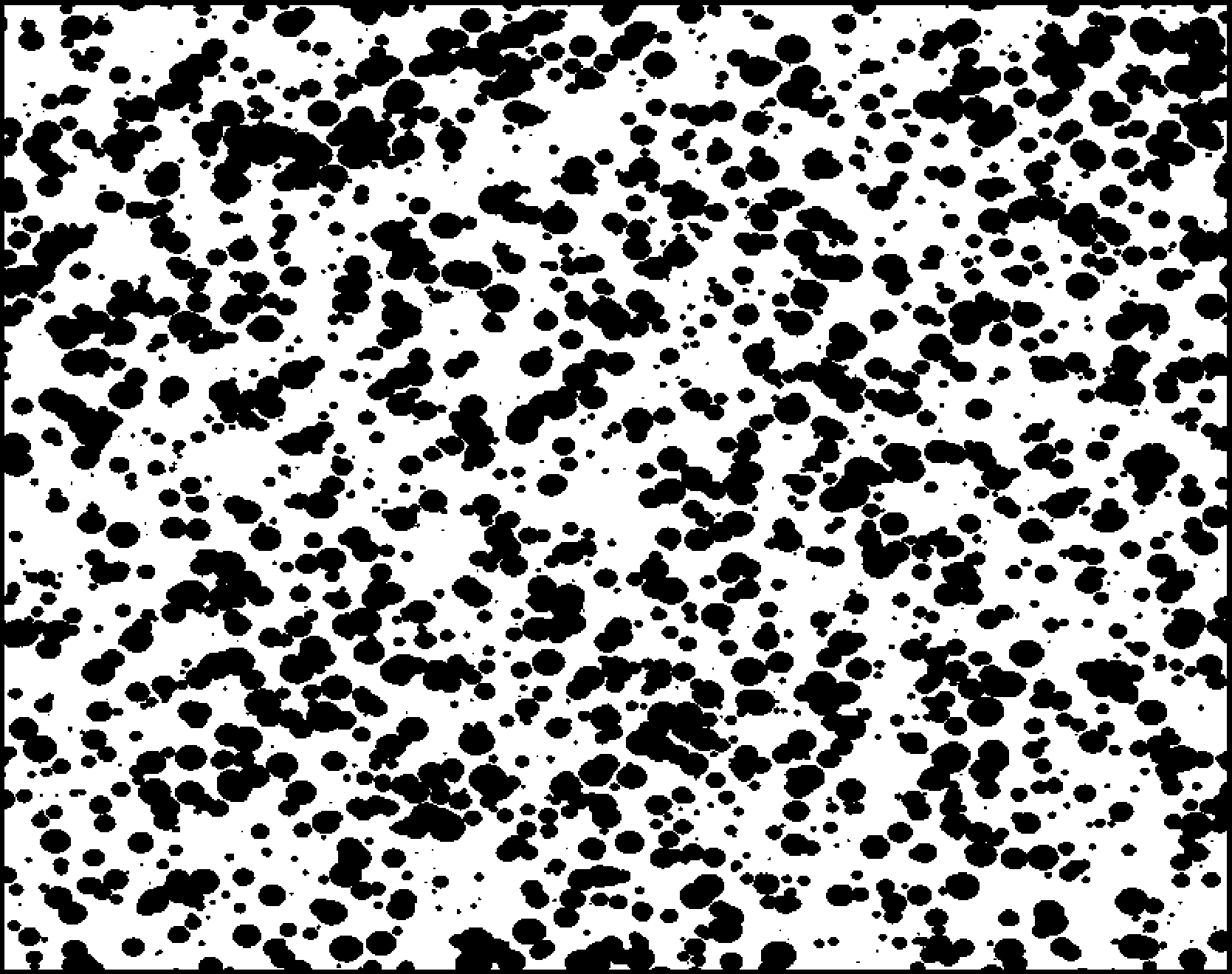}
\\[5mm]
\includegraphics[clip, width=.22\textwidth]{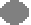}
\includegraphics[clip, width=.22\textwidth]{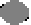}
\includegraphics[clip, width=.22\textwidth]{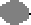}
\includegraphics[clip, width=.22\textwidth]{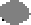}
\caption{We consider two pictures of a simulated birth--and--growth process, at two different time instants, that in our notations are $X$ and $Y$. Emphasizing the differences, we report here the magnified pictures of the true growth used for the simulation, the computed $\tildeSpeed{2}{W}$, $\tildeSpeed{1}{W}$ and $\tildeSpeed{1}{W\ominus\check{\cpt}}$. Note that they agree with Proposition \ref{pro: properties tildeSpeed} and Proposition \ref{pro: G2 consistent as W expand to X} since $\tildeSpeed{1}{W\ominus\check{\cpt}} \supseteq \tildeSpeed{1}{W} \supseteq \tildeSpeed{2}{W}$.}
\label{fig: Matlab4}
\end{center}
\end{figure}

\subparagraph{\textbf{A General Definition of
$\tildeSpeed{2}{W}$.}} The following proposition shows that the
estimator in \eqref{eq: G2(W,K)K} can be defined in an equivalent
way by
$$
\tildeSpeed{2}{W}(Z)=\graffa{\quadra{ \cut{Y}{W} \cup \add{Z} } \ominus \cut{\check{X}}{W}}_{\cpt};
$$
where $\add{X}$ in \eqref{eq: G2(W,K)K} is substituted by $\add{Z}$ with
\begin{equation}\label{eq: B_(W setminus W+K)<C<W}
\cut{X}{W\setminus \tonda{W\ominus \check{\cpt}}}\subseteq Z\subseteq W.
\end{equation}
In other words, we are saying that, under condition \eqref{eq:
B_(W setminus W+K)<C<W}, $\tildeSpeed{2}{W}(Z)$ does not depend on
$Z$. From a computational point of view, this means that $Z$ can
be chosen in a way that reduces the computational costs.
On the one hand, the best choice of $Z$ seems to be the smallest possible set, i.e.\ $Z=\cut{X}{W\setminus \tonda{W\ominus \check{\cpt}}}$. On the other hand, in order to get $\cut{X}{W\setminus \tonda{W\ominus \check{\cpt}}}$, we have to compute $\tonda{W\ominus \check{\cpt}}$ that may be costly if at least one between $W$ and $\cpt$ has a ``bad shape'' (for instance it is not a rectangular one).
\begin{pro}
If $Z_1,Z_2\in\Parts{\Banach}$ both satisfy condition \eqref{eq: B_(W
setminus W+K)<C<W}, then $\tildeSpeed{2}{W}(Z_1)=\tildeSpeed{2}{W}(Z_2)$.
\end{pro}
\begin{myproof}
It is sufficient to prove:
\begin{mynumerate}
\mynumber $Z_1\subseteq Z_2$ implies $\tildeSpeed{2}{W}(Z_1)\subseteq
\tildeSpeed{2}{W}(Z_2)$;
\mynumber $\tildeSpeed{2}{W}(W)\subseteq
\tildeSpeed{2}{W}\tonda{\cut{X}{W\setminus \tonda{W\ominus \check{\cpt}}}}$.
\end{mynumerate}

In fact, (1) and (2) imply that $\tildeSpeed{2}{W}(W) =
\tildeSpeed{2}{W}\tonda{\cut{X}{W\setminus \tonda{W\ominus \check{\cpt}}}}$. At
the same time they imply $\tildeSpeed{2}{W}(Z) =
\tildeSpeed{2}{W}\tonda{\cut{X}{W\setminus \tonda{W\ominus \check{\cpt}}}}$
holds for every $Z$ that satisfies \eqref{eq: B_(W setminus W+K)<C<W};
that is the thesis.
\\%
\emph{STEP (1)} is a consequence of the following implications
\begin{eqnarray*}
Z_1\subseteq Z_2 &\Rightarrow& Z_1\Minkowski\cpt \subseteq Z_2\Minkowski\cpt,
\\
&\Rightarrow& \cut{Y}{W}\cup \quadra{\tonda{Z_1\Minkowski\cpt}\setminus W}
\subseteq \cut{Y}{W}\cup \quadra{\tonda{Z_2\Minkowski\cpt}\setminus W},
\\
&\Rightarrow& \tildeSpeed{2}{W}(Z_1)\subseteq \tildeSpeed{2}{W}(Z_2);
\end{eqnarray*}
where the last one holds since $X_1\ominus Y \subseteq X_2\ominus Y$ if
$X_1\subseteq X_2$ (see \cite{ser84}).

Before proving the second step, we show that $\tildeSpeed{2}{W}\tonda{Z} = \tildeSpeed{2}{W}\tonda{\cut{Z}{W\setminus \tonda{W\ominus\check{\cpt}}}}$ for all $Z$ that satisfies \eqref{eq: B_(W setminus W+K)<C<W}. This statement is true if $\tonda{\cut{Z}{W\setminus \tonda{W\ominus \check{\cpt}}}\Minkowski\cpt}\setminus W$ and $\tonda{Z\Minkowski\cpt}\setminus W$ are the same set. Since Minkowski sum is distributive with respect to union, we get
\begin{eqnarray*}
\tonda{Z\Minkowski\cpt}\setminus W &=& \quadra{\tonda{ \cut{Z}{W\setminus
\tonda{W\ominus \check{\cpt}}}\cup \cut{Z}{W\ominus \check{\cpt}} }\Minkowski\cpt }\setminus W
\\
&=& \quadra{\tonda{ \cut{Z}{W\setminus \tonda{W\ominus \check{\cpt}}} \Minkowski\cpt }\setminus W} \cup \quadra{\tonda{ \cut{Z}{W\ominus \check{\cpt}}
\Minkowski\cpt }\setminus W}.
\end{eqnarray*}
Then we have to prove that $\quadra{\tonda{ \cut{Z}{W\ominus \check{\cpt}}
\Minkowski\cpt }\setminus W}=\emptyset$ :
\begin{eqnarray*}
\tonda{ \cut{Z}{W\ominus \check{\cpt}} \Minkowski\cpt }\setminus W
&=& \graffa{\quadra{Z\cap \tonda{W\ominus \check{\cpt}}} \Minkowski\cpt
}\setminus W
\\
&\subseteq& \graffa{\tonda{Z\Minkowski\cpt} \cap \quadra{ \tonda{W\ominus
\check{\cpt}} \Minkowski\cpt } }\setminus W
\\
&\subseteq&  \quadra{\tonda{Z\Minkowski\cpt} \cap W}\setminus W =\emptyset.
\end{eqnarray*}
\emph{STEP (2)}. Since $\tildeSpeed{2}{W} \tonda{\cut{X}{W}} = \tildeSpeed{2}{W}\tonda{\cut{X}{W\setminus \tonda{W\ominus \check{\cpt}}}}$, thesis becomes $\tildeSpeed{2}{W}(W) \subseteq \tildeSpeed{2}{W}(\cut{X}{W})$.
Let $g\in\tildeSpeed{2}{W}(W)$. We must prove
$g\in\tildeSpeed{2}{W}(\cut{X}{W})$, i.e.\ for every $x\in \cut{X}{W}$
$$
g+x \in \cut{Y}{W}, \qquad \textrm{ or }\qquad g+x \in
\tonda{\cut{X}{W} \Minkowski\cpt }\setminus W.
$$
Since $g\in\tildeSpeed{2}{W}(W)$, for any $x\in\cut{X}{W}$ we can have two
possibilities
\myitem[(a)] $g+x\in \cut{Y}{W}$,
\myitem[(b)] $g+x\in \tonda{W\Minkowski\cpt}\setminus W$.
\\%
It remains to prove that (b) implies $g+x\in \tonda{\cut{X}{W}\Minkowski\cpt}\setminus W$.
In particular, (b) implies $g+x\in W^C$. At the same time $g+x \in \cut{X}{W} \Minkowski\cpt$,  i.e.\ $g+x\in \tonda{\cut{X}{W} \Minkowski\cpt}\setminus W$.
\end{myproof}

\section{Hitting Function Associated to $\nucleation$}\label{sec: capacity of nucleation}

In many practical cases, an observer, through a window $W$ and at two different instants, observes the nucleation and growth processes namely $X$ and $Y$. According to Section \ref{sec: estimators of G} we can estimate $\speed$ via the consistent estimator $\tildeSpeed{2}{W}$ or $\tildeSpeed{1}{W}$ (in the following we shall write $\tildeSpeed{}{W}$ meaning one of them).
From the birth--and--growth process point of view, it is also interesting to test whenever the nucleation process $\nucleation=\graffa{\nucleation_n}_{n\in\bbN}$ is a specific \racs{} (for example a Boolean model or a point process).
In general, we cannot directly observe the $n$--th nucleation $\nucleation_n$ since it can be overlapped by other nuclei or by their evolutions. Nevertheless, we shall infer on the hitting function associated to the nucleation process $T_{\nucleation_n}(\cdot)$.
Let us consider the decomposition given by \eqref{eq: decomposition} $ Y=(X \Minkowski \Dspeed) \cup \nucleation $ then the following proposition is a consequence of Remark \ref{oss: capacity functional of XUY}.
\begin{pro}\label{pro: capacity of nucleation}
If $(\speed,\nucleation)$ is a $X$--decomposition of $Y$ such that $\nucleation$ is independent on $X$ and on $\speed$, then, for each $\cmpt\in\eClosed[k]{\Banach}$, $$
T_{Y}\tonda{\cmpt} = T_{X\Minkowski\speed} \tonda{\cmpt} +
T_{\nucleation}\tonda{\cmpt} - T_{X\Minkowski\speed} \tonda{\cmpt}
T_{\nucleation}\tonda{\cmpt},
$$
that, in terms of $Q_{\cdot}(\cmpt)= \tonda{1-T_{\cdot}\tonda{\cmpt}} $, is equivalent to
$$
Q_{Y}(\cmpt) = Q_\nucleation(\cmpt)Q_{X\Minkowski \speed}(\cmpt).
$$
\end{pro}
In other words, the probability for the exploring set $\cmpt$ to miss $Y$ is the probability for $\cmpt$ to miss $\nucleation$ multiplied by the probability for $\cmpt$ to miss $X\Minkowski\speed$.

\begin{oss}
Working with data we shall consider two estimators of the hitting function (we refer to \cite[p. 57--63]{mol97} and references therein). In particular, if $X$ is a stationary ergodic \racs{}, then $T_{X}(\cdot)$ can be estimated by a single realization of $X$ and two empirical estimators are given by
$$
\widehat{T}_{X,W}(\cmpt) = \frac{\leb\tonda{
\tonda{X\Minkowski\check{\cmpt}}\cap\tonda{W\ominus \cmpt_0} }}{
\leb\tonda{W\ominus \cmpt_0}},\quad \cmpt\in\eClosed[k]{\Banach};
$$
where $\leb$ is the Lebesgue measure on $\Banach=\bbR^d$ and $\cmpt_0$ is a compact set such that $\cmpt\subset \cmpt_0$ for all $\cmpt\in\eClosed[k]{\Banach}$ of interest.
\end{oss}
A \emph{regular closed} set in $\Banach$ is a closed set $\speed\in\Closed{\Banach}$ for which $\speed=\closure{\interior{\speed}}$; i.e.\ $\speed$ is the closure (in $\Banach$) of its interior.
\begin{pro}\label{pro: regular closed set}
Let $\speed\in\Closed[k]{\Banach}$ be a regular closed subset in $\Banach$. Then, for every $X\in\Closed{\Banach}$, $X\Minkowski \speed$ is a regular closed set.
\end{pro}
\begin{myproof}
Since $X\Minkowski \speed$ is a closed set, $\closure{\interior{(X\Minkowski \speed)}}\subseteq X\Minkowski \speed$. It remains to prove that $X\Minkowski \speed\subseteq \closure{\interior{(X\Minkowski \speed)}}$. Let $y\in X\Minkowski \speed$, then there exists $x\in X$ and $g\in \speed$ such that $y=x+g$. If $g\in\interior{\speed}$, then there exists an open neighborhood of $g$ for which $U(g)\subseteq \interior{\speed}$ and $x\Minkowski U(g)$ is an open neighborhood of $x+g$ included in $X\Minkowski \speed$; i.e.\ $x+g\in\interior{(X\Minkowski \speed)}$. On the other hand, let $g\in\partial \speed=\speed\setminus \interior{\speed}$, then there exists $\graffa{g_n}_{n\in\bbN}\subset \speed$ such that $g_n\rightarrow g$ and $g_n\in\interior{\speed}$, for all $n\in\bbN$. 
Thus, for every $n\in\bbN$, $x+g_n$ is an interior point of $X\Minkowski \speed$ and $x+g_n \rightarrow x+g\in \closure{\interior{(X\Minkowski \speed)}}$.
\end{myproof}

\begin{pro}[{(See \cite[Theorem 4.5 p. 61]{mol97} and references therein)}]
Let $X$ be an ergodic stationary random closed set. If the random set $X$ is almost surely regular closed
\begin{equation}\label{eq: choquet estimation uniform convergence}
\supKzero \modulo{\widehat{T}_{X,W}(\cmpt)-T_X(\cmpt)}\rightarrow 0, \quad \textrm{a.s.}
\end{equation}
as $W\uparrow\Banach$ and for every $\cmpt_0\in\Closed{\Banach}$.
\end{pro}

\begin{oss}
Proposition \ref{pro: regular closed set}, together to Equation \eqref{eq: discrete set process} means that, if $\graffa{\speed_n}_{n\in\bbN}$ is a sequence of almost surely regular closed sets, then $\graffa{\Theta_n}_{n\in\bbN}$ is so.
\end{oss}
The following Theorem shows that the hitting functional ${Q}_{\nucleation}$
of the hidden nucleation process can be exstimated by the observable quantity
$\widetilde{Q}_{\nucleation,W}$, where for every $\cmpt\in\eClosed[k]{\Banach}$,
\begin{equation}\label{eq:defin_functional}
\widetilde{Q}_{\nucleation,W}(\cmpt) : = \frac{\widehat{Q}_{Y,W}(\cmpt)}{\widehat{Q}_{X\Minkowski \tildeSpeed{}{\textrm{\tiny $W$}},W}(\cmpt) },
\end{equation}
and $\tildeSpeed{}{W}$ is given by \eqref{eq: G1(W,K)K} or \eqref{eq: G2(W,K)K}.
\begin{teo}
Let $X,Y$ be two \racs{} a.s. regular closed. Let $(\speed,\nucleation)$ be a $X$--decomposition of $Y$ with $\nucleation$ a stationary ergodic \racs{} independent  on $\speed$ and $X$.
Assume that $\speed$ is an a.s. regular closed set and $\widetilde{Q}_{\nucleation,W}$
defined in Equation~\eqref{eq:defin_functional}. Then, for any $\cmpt\in\eClosed[k]{\Banach}$,
$$
\modulo{\widetilde{Q}_{\nucleation,W}(\cmpt)-Q_{\nucleation}(\cmpt)}
\mathop{\longrightarrow}_{W\uparrow\Banach} 0,
\quad \textrm{a.s.}
$$
\end{teo}
\begin{myproof}
Let $\cmpt\in\eClosed[k]{\Banach}$ be fixed.
For the sake of simplicity, $Q_\cdot$, $\widetilde{Q}_\cdot$ and $\widehat{Q}_\cdot$ denote
$Q_\cdot(\cmpt)$, $\widetilde{Q}_{\cdot,W}(\cmpt)$ and $\widehat{Q}_{\cdot,W}(\cmpt)$
respectively. Thus,
$$
\modulo{\widetilde{Q}_{\nucleation}-Q_\nucleation} = \modulo{\frac{\widehat{Q}_{Y}}{\widehat{Q}_{X\Minkowski \tildeSpeed{}{W}}}- \frac{Q_{Y}}{Q_{X\Minkowski \speed}}}
= \modulo{\frac{\widehat{Q}_{Y}Q_{X\Minkowski \speed} - Q_{Y}\widehat{Q}_{X\Minkowski \tildeSpeed{}{W}}}{ \widehat{Q}_{X\Minkowski \tildeSpeed{}{W}}Q_{X\Minkowski \speed}} }.
$$
Since $Y\supseteq X\Minkowski \tildeSpeed{}{W}$, $\widehat{Q}_{X\Minkowski \tildeSpeed{}{W}} > \widehat{Q}_Y$. Accordingly to \eqref{eq: choquet estimation uniform convergence}, $\widehat{Q}_Y $ converges to $Q_Y$ that is a positive quantity. Thus, thesis is equivalent to prove that
$$
\modulo{\widehat{Q}_{Y}Q_{X\Minkowski \speed} - Q_{Y}\widehat{Q}_{X\Minkowski \tildeSpeed{}{W}}} \rightarrow 0, \quad \textrm{a.s.}
$$
as $W\uparrow \Banach$. The following inequalities hold
\begin{eqnarray*}
\modulo{\widehat{Q}_{Y}Q_{X\Minkowski \speed} - Q_{Y}\widehat{Q}_{X\Minkowski \tildeSpeed{}{W}}} &\le& Q_{X\Minkowski \speed}\modulo{\widehat{Q}_{Y}-Q_Y} + Q_Y\modulo{Q_{X\Minkowski \speed}-\widehat{Q}_{X\Minkowski \tildeSpeed{}{W}}}
\\%
&\le& Q_{X\Minkowski \speed}\modulo{\widehat{Q}_{Y}-Q_Y} +
\\%
&& Q_Y\modulo{Q_{X\Minkowski \speed}-Q_{X\Minkowski \tildeSpeed{}{W}}} + Q_Y\modulo{Q_{X\Minkowski \tildeSpeed{}{W}}-\widehat{Q}_{X\Minkowski \tildeSpeed{}{W}}}.
\end{eqnarray*}
Proposition \ref{pro: X+Y stationary} and Proposition \ref{pro: regular closed set} guarantee that $X\Minkowski \speed$ is a stationary ergodic \racs{} and a.s. regular closed, then we can apply \eqref{eq: choquet estimation uniform convergence} to the first and the third addends. It remains to prove that
\begin{equation}\label{eq:proof_dist}
 \modulo{Q_{X\Minkowski \speed}-Q_{X\Minkowski \tildeSpeed{}{W}}}
\rightarrow 0\quad\textrm{as }W\uparrow\Banach.
\end{equation}
Since Minkowski sum is a continuous map from $\eClosed{\Banach}\times\eClosed[k]{\Banach}$ to $\eClosed{\Banach}$ (see \cite{ser84}), $\tildeSpeed{}{W}\downarrow\speed$ a.s. implies $X\Minkowski\tildeSpeed{}{W} \downarrow X\Minkowski\speed$ a.s.\@
As a consequence, we get that $X\Minkowski\tildeSpeed{}{W} \downarrow X\Minkowski\speed$
in distribution \cite[p.~$182$]{ngu06}, which is Equation~\eqref{eq:proof_dist}.
%
\end{myproof}

%
%
%
%

\end{document}